\documentstyle[preprint,aps,prd]{revtex}

\begin{document}

\tighten

\title {Classicality of the order parameter during a phase transition}    

\author{Fernando C.\ Lombardo$^{1,2}$  \footnote{Electronic address: 
f.lombardo@ic.ac.uk}, Francisco D.\ Mazzitelli $^1$ 
\footnote{Electronic address: fmazzi@df.uba.ar}, and Diana Monteoliva 
$^1$
\footnote{Electronic address: monteoli@df.uba.ar}}

\address{{\it
$^1$ Departamento de F\'\i sica, Facultad de Ciencias Exactas y Naturales\\ 
Universidad de Buenos Aires - Ciudad Universitaria, 
Pabell\' on I\\ 
1428 Buenos Aires, Argentina\\
$^2$ Blackett Laboratory, Imperial College, London SW7 2BZ}} 

\maketitle

\begin{abstract}
We analyze the quantum to classical transition of the order parameter
in second order phase transitions. We consider several toy models in
non relativistic quantum mechanics.  We study
the dynamical evolution of a wave packet initially peaked around a local
maximum of the potential using variational approximations and also
exact numerical results. The influence of the 
environment on the evolution of the density matrix and the Wigner 
function is analyzed in great detail. We also discuss the relevance of
our results  to the analysis of phase transitions in field theory. 
In particular,
we argue that previous results about classicality of the order parameter
in $O(N)$ models may be consequences of the large $N$ approximation.
\end{abstract}

\vskip 2cm
PACS numbers: 05.70.Fh,  03.70.+k, 05.40.Jc
\newpage
\section{Introduction}

The emergence of classical behaviour from a quantum system is 
a problem of interest in many branches 
of physics \cite{giulmol}. As is  
well known, the quantum to classical transition involves two necessary 
and related conditions: {\it correlations}, i.e. the 
Wigner function of a quantum system should have a peak at the classical 
trajectories; and {\it decoherence}, that is, there should be no 
interference between classical trajectories. To
study quantitatively 
the emergence of classicality, it is essential to consider the
interaction of the system with its environment, since both the 
loss of quantum coherence
and the onset of classical correlations depend 
strongly on this interaction \cite{unruhzu}.
Using this point of view, classicality is an emergent property of an
open quantum system. The strength of the coupling between system and 
environment 
sets the 
decoherence time which, roughly speaking, indicates the timescale after
which
the system can be considered as classical. 

Our concern in this paper will be the analysis of the quantum to classical
transition of the order parameter during second order phase transitions, in 
which the effective potential has a local maximum.
In condensed matter physics, there are several systems in which the dynamics
of phase transitions can be studied experimentally (superfluids for example)
 \cite{zurek}.
In the standard big-bang cosmological model, phase transitions occur
at the GUT and EW scales. During these phase transitions topological 
defects are inevitable, and they may have played a fundamental role in the 
formation of large scale structure (strings) \cite{zurek}. Moreover, 
superabundance of some
topological defects may contradict observational evidence (magnetic 
monopoles). 
In order 
to solve this and related problems it is widely accepted that, before the 
radiation dominated era, the Universe expanded exponentially (inflationary 
period). This exponential expansion takes place during a second order phase
transition.

In all the above mentioned examples there is an order parameter 
which evolves
from the false to the true vacuum of the theory: the 
Higgs fields in GUT and EW phase transitions, the inflaton 
field(s) in inflationary
models, etc. 
Although these are quantum scalar fields with vanishing mean value
(due to the symmetry of the initial quantum state), the order parameter is 
usually 
treated as a classical object. The
classical behaviour is fundamental to define and count the topological
defects \cite{rivers}, and to justify the fact that 
some gauge and fermion fields aquire masses.

In the present work, we will analyze
the classicality of the order parameter during a second order phase 
transition.
Of course, in a realistic model one should address this problem in the 
context of  quantum field theory. In fact, 
a possible approach would be to follow the analysis started by two 
of us in Ref.\cite{nos}, where we studied
the emergence of classical
inhomogeneities from quantum fluctuations for a self-interacting quantum 
scalar field. We have investigated there
the decoherence induced on the long-wavelength field modes
by coarse graining the field modes with wavelength shorter than a critical
value, in order to show how the system becomes classical due to the 
interaction with its environment (in that case composed by the 
short-wavelength field modes of the same field). The 
classicality of the order parameter could be analyzed along the same lines
by considering a model with 
spontaneous symmetry breaking.

This is an extremely difficult problem because, as has been 
pointed out in the literature, and as we will stress in what 
follows, non perturbative and non Gaussian effects are relevant
in the quantum to classical transition of the order parameter. 
In field theory, 
it is very difficult to go beyond 
perturbative or mean field methods (Hartree, $1/N$, etc).
For this reason, in the present
paper we will be mainly concerned with toy models: closed and open 
systems \cite{hpz} in non relativistic
quantum mechanics. 
We will study the spread of a wave packet initially centered
around the local maximum of a double well potential, paying 
particular attention to the influence of the environment on
the Wigner function and on the reduced density matrix.
We will also
discuss 
the relevance of our results to the analysis of the field theory
phase transitions.

The paper is organized as follows. In the next section we review the approach
of Guth and Pi to describe the initial stages of the quantum dynamics of
the phase transition using an inverted harmonic oscillator. We consider the
upside down harmonic oscillator with and without environment, in order to
emphazise the relevance of the environment in the quantum to classical 
transition. 
In section III we study  the evolution of a wave packet initially
centered on the top of a double well potential. We describe a Gaussian 
variational
calculation and an improved version of it. We find that, in the Gaussian 
approximation, the Wigner function is positive for all times but it does
not describe classical correlations unless the system is coupled to 
an environment. The Gaussian approach breaks down as the wave packet spreads 
out and
explores the minimum of the potential. The improved variational
approximation describes the 
dynamics of the
system beyond that point. However, as the wave function is not Gaussian,
the Wigner function is no longer positive.
In section IV we describe the exact numerical evaluation of the evolution
of the wave packet. We show that, as the coupling between the system and 
the environment increases, the decoherence time decreases. Due to
the nonlinearities of the potential, when the coupling vanishes there is
no classical limit, not even classical correlations. In section V
we analyze previous works on the quantum to classical transition
in field theory in the light of the results for the toy models. 

\section{The inverted harmonic oscillator: role of the environment} 

In a cosmological scenario, at very high temperatures
the effective potential for a scalar field $\phi$ has a minimum at 
$\phi =0$. As the temperature decreases this minimum  
becomes unstable, and the stable minima move to a nonvanishing
value of the field. During the phase transition, the system evolves
to its true vacuum.

A sudden quench phase transition can be described by a field theory
in which there is an instantaneous change of sign in the mass term of the 
scalar field 
\begin{equation} S[\phi] = \int d^4x \big[{1\over{2}}\partial_\mu\phi
\partial^\mu \phi - {1\over{2}}m(t)^2 \phi^2 - {1\over{4!}}\lambda 
\phi^4\big],\label{1}
\end{equation} 
where  $m^2(t) = \mu^2 > 0$  for $t < t_0$ and 
 $m^2(t)= - \mu^2$  for $t > t_0$ (we will take 
$t_0 = 0$ for simplicity). This change of sign   
in $m^2(t)$ breaks the global symmetry for $t>0$.

Guth and Pi \cite{guthpi} considered an upside down harmonic oscillator as a 
toy model 
to describe the quantum behaviour of this unstable 
system
\begin{equation}
S[x]= \int_0^t ds {1\over{2}} M ({\dot x}^2 + \Omega_0^2 x^2).
\label{invosc}
\end{equation}
This toy model 
should be a good approximation for the early time evolution of the 
phase transition, as long as one can neglect the non-linearities
of the potential.

If the initial  wave function is Gaussian, it will remain Gaussian
for all times  
(with time dependent parameters that set its amplitude and spread).
The density matrix will be of the form
\begin{equation} \rho (x, x',t) = N(t) e^{- A(t) x^2 - B(t) x'^2},
\label{rhogauss}\end{equation}
where $N(t)$ is a time-dependent real normalization function; $A(t)$ and 
$B(t)$ are time dependent complex coefficients, which satisfy $A= a + i b 
= B^*$. Using new variables $\Sigma = (x+ x')/2$ and $\Delta = (x - x')/2$, 
the density matrix can be re-written as
\begin{equation} \rho (t,\Sigma ,\Delta) = N(t) e^{-2 a \Delta^2} 
e^{-2 a \Sigma^2} e^{-4 
i b \Sigma\Delta}.\end{equation}
For such a density matrix, the asociated Wigner function is also
Gaussian and can be exactly 
evaluated as (here and what follows we  set $\hbar \equiv 1$)
\begin{equation}W(x,p,t )={1\over{2\pi}} \int_{-\infty}^{+\infty}
dy e^{i p y} \rho (x+{y\over{2}}, x-{y\over{2}}) = 
{1\over{\pi}} e^{-2 a x^2} e^{-{(p - 2 x b)
^2\over{2a}}}.
\end{equation}

The coefficient $2a$ gives the spread of the Wigner distribution around the 
classical trajectory, and, at the same time, 
$(2a)^{-1}$ measures the 
importance 
of the non-diagonal terms in the density matrix. 
As is well known, there is a compromise between the spread of the Wigner 
function and the diagonalization of the density matrix: as one 
becomes peaked the other becomes non diagonal \cite{laflamme}.

In Fig. 1, we  show
the time dependence of $2a$ obtained from the Schr\"oedinger 
equation. It is an exponentially  decreasing function. 
Therefore, although 
the density matrix is non-diagonal, the Wigner distribution 
becomes peaked around the classical trajectory for long 
times. The unstable quantum evolution shows classical behaviour,
in the sense that one can obtain a classical 
probability distribution for the unstable particle \cite{guthpi}. Strictly 
speaking,
we do not have classical limit but ``classical
correlations'', because the density matrix is not diagonal. As has been
recently pointed out by Kiefer et al \cite{kiefer}, 
this may be enough for the quantum to classical
transition of free field fluctuations in a cosmological setting. However,
as we will see in the next sections, this is not true for the double well
potential.

The classical limit exhibited by the quantum particle in the upside down 
potential requires the  coefficient $2a$ to reach its minimum value. 
This limit 
is obviously reached at large times, when the particle is far away 
from the potential top. The ``correlation time" depends
on the parameters of the potential.  As we will now see, the presence of an
environment changes drastically this situation.

Let us consider the
unstable quantum particle (characterized by its mass $M$ and 
its bare frequency $\Omega_0$) linearly coupled to an environment 
composed by an infinite set of harmonic oscillators (of mass $m_n$ and 
frequency $\omega_n$). We may write the total action corresponding to the 
system-environment model as
\begin{eqnarray}S[x,q_n] &=& S[x] + S[q_n] + S_{\rm int}[x,q_n]\nonumber \\
&=& \int_0^t ds \left[{1\over{2}} M ({\dot x}^2 + \Omega_0^2 x^2) + 
\sum_n {1\over{2}} m_n ({\dot q}_n^2 - \omega_n^2 q_n^2)\right] - 
\sum_n C_n x q_n,\end{eqnarray}  
where $x$ and $q_n$ are the coordinates of the particle and the oscillators, 
respectively. The unstable particle is coupled linearly to each oscillator 
with strength $C_n$.

The relevant objects to analize the quantum to classical transition in this
model are the reduced density matrix, and the associated Wigner function
\begin{eqnarray}
\rho_{\rm r} (x,x',t)&=& \int dq_n\,\, \rho (x,q_n,x',q_n,t)\nonumber\\
W_{\rm r} (x,p,t)&=& {1\over{2\pi}} \int_{-\infty}^{+\infty} dy~ 
e^{ipy} ~ \rho_{\rm r}(x+{y\over{2}}, x-{y\over{2}},t).
\end{eqnarray} 
The reduced density matrix satisfies a 
master equation.
Hu-Paz-Zhang \cite{hpz} have evaluated the master equation for the quantum 
Brownian motion problem. 
Following the same procedure, we can
write the master equation for 
the unstable particle \cite{pazzurek}. As 
the coupling between system and environment is lineal, the result is exact, 
and can be easily obtained it replacing $\Omega_0$ by $i \Omega_0$ in the 
Hu-Paz-Zhang result
\begin{equation}   \dot{\rho}_{\rm r} =-i\bigl[{H}_{\rm 
syst}-{1\over{2}}M{\tilde \Omega}^2(t)
,\rho_{\rm r}\bigr]
+2i\gamma(t)\bigl[x,\bigl\{ p,\rho_{\rm r}\bigr\}\bigr]
-D(t)\bigl[x,\bigl[ x,\rho_{\rm r}\bigr]\bigr] 
 -f(t)\bigl[x,\bigl[ p,\rho_{\rm r}\bigr]\bigr].
\label{master}\end{equation}
The time dependent coefficients are given by
\begin{eqnarray}
\tilde\Omega^2(t) &=& -{2\over M}\int_0^t dt'\cosh(\Omega_0 t') 
\eta(t')\nonumber \\
\gamma(t) &=& -{1\over 2M\Omega_0}\int_0^t dt'\sinh(\Omega_0 t') 
\eta(t')\nonumber 
\\
D(t) &=& \int_0^t dt'\cosh(\Omega_0 t') \nu(t')\label{coef} \\
f(t) &=& -{1\over M\Omega_0}\int_0^t dt'\sinh(\Omega_0 t') \eta(t'),\nonumber
\end{eqnarray}
${\tilde \Omega}(t)$ renormalizes the natural frequency of the 
particle, $\gamma (t)$ is the dissipation coefficient, and $D(t)$ and 
$f(t)$ are 
the difusion coefficients, which produce the decoherence effects.
$\eta (t)$ and $\nu (t)$ are the dissipation and noise kernels, 
respectively, 
\begin{eqnarray}\eta (t)& =& \int_0^\infty d\omega I(\omega ) \sin \omega t 
\nonumber \\
\nu (t) &=& \int_0^\infty d\omega I(\omega ) \coth {\beta
\omega\over{2}} \cos \omega t\nonumber, \nonumber\end{eqnarray} 
and $I(\omega )$ is the spectral density of the environment.
In the high temperature limit of an ohmic environment (where $I(\omega ) 
\propto \omega $) the coefficients in Eq.(\ref{coef}) become
constants. In particular, the diffusion coefficient can be
approximated by $D \simeq 2 \gamma_0 k_{\rm B} T M$, where $\gamma_0$ is 
the dissipation coefficient \cite{hpz}. In this limit, while 
$\gamma_0$ is a constant and $D \propto T$, the dissipation coefficient 
is
$f \propto T^{-1}$. Therefore the term proportional to 
$D$ is the relevant term in the master equation. 

Alternatively, one can write an equation of the Fokker-Planck type
for the reduced Wigner function \cite{hpz}. It is given by
\begin{equation}\dot{W_r}(x,p,t)= -\{H_{\rm syst}- {1\over{2}} M 
{\tilde\Omega}^2(t),W_r\}_{\rm PB} +
 2 \gamma (t) \partial_p(pW_r)
+ D(t) \partial^2_{pp}W_r - f(t) \partial^2_{px}W_r,
\label{fokker}\end{equation}
where the first term on the right-hand side is a Poisson bracket.

Let us solve Eqs.(\ref{master}) and (\ref{fokker}) using again a Gaussian 
ansatz for the reduced density matrix. The adequate generalization of
Eq. (\ref{rhogauss}) is
\begin{equation}\rho_r (x,x',t) = N(t) e^{- A(t) x^2 - B(t) x'^2 - 
C(t) x x'},
\label{rhogauss2}
\end{equation}
where $C(t)$ is a real function.
The master equation, in the high temperature limit, becomes
\begin{eqnarray}\dot{a} &=& 4 a b - 2 \gamma_0 a + 2 \gamma_0 {\tilde T} +
 \gamma_0 C
\nonumber \\
\dot{b} &=& - 2 a^2 + 2 b^2 - 2 \gamma_0 b + {1\over{2}} C^2 - {1\over{2}} 
{\tilde \Omega}^2 \nonumber \\
\dot{C} &=& 4 \gamma_0 a + 4 C b - 2 \gamma_0 C - 4 \gamma_0 {\tilde T}
 \nonumber \\
\dot{N} &=& 2 N b, \label{coopereqs}\end{eqnarray}
where we are denoting ${\tilde T} = k_{\rm B} T$, and we have set $M = 1$.

Using again the variables $\Sigma$ and $\Delta$ the reduced density
matrix reads
\begin{equation}\rho_r (\Sigma ,\Delta ,t) = N(t) e^{-(2 a - C) \Delta^2} 
e^{- (2 a + C) 
\Sigma^2} e^{- 4 i b \Sigma \Delta},\end{equation}
while the reduced Wigner function is exactly evaluated as
\begin{equation}W_r(x,p,t) = {1\over{\pi}} \sqrt{{2 a + C\over{2 a - C}}}
e^{-( 2 a + C) x^2}
e^{-{(p - 2 x b)^2\over{(2 a - C)}}},\end{equation}

From the last two equations we see that the relevant function to 
describe correlations and decoherence is now $2a-C$. For $2a - C =O(1)$
we have both correlations and decoherence. The set of Eqs.(\ref{coopereqs}) 
can be easily solved numerically. In Fig. 2 we show 
the behaviour of $2a-C$ as a function of time. We see that it tends
asymptotically to a constant of order one (of course the asymptotic
value depends on the properties of the environment).

The main conclusion of this section is the following.
In order to study a sudden quench phase transition, at early times 
we can use the upside down potential. When the system is isolated,
due to the high squeezing of the initial wave packet $x$ and $p$ become
classically correlated. The density matrix is not diagonal. The 
``correlation time" depends on the shape of the potential. 
When the particle is coupled to an environment, a true quantum to classical
transition takes place. The Wigner function becomes peaked around a 
classical trajectory and the density matrix diagonalizes. The 
decoherence time depend on the diffusion coefficient $D$.   

\section{Including self-interaction: variational approximations}

Let us now consider a more realistic model by adding 
a $\lambda x^4$ term to the Lagrangian  of the unstable 
quantum particle. 

Of course the problem no longer admits an exact, analytical solution.
Before presenting the numerical solution (Section IV), we would like to 
discuss some analytical approximations based on the time
dependent variational method developed by Jackiw and Kerman \cite{jackiwk}. 
This
will be useful to understand the validity of similar
approximations in field theory.

The variational method is based on the definition of an ``effective action'' 
\begin{equation} \Gamma = \int dt \langle \psi\vert i 
{\partial\over{\partial 
t}} - {\hat H}\vert \psi \rangle .\end{equation}
When $\Gamma$ is stationary against variations of the 
state $\vert \psi\rangle$ (with $\langle \psi\vert \psi\rangle = 1$), 
the state satisfies the Schr\"oedinger equation.
Approximated solutions are obtained by minimizing the effective action
within a family of trial wave functions. 
 
Following the work of Jackiw and Kerman, Cooper et al \cite{cooper} studied 
the dynamics of a quantum particle in a double well potential
in the so called Gaussian approximation. 
In this approximation, the problem is equivalent to the inverted
oscillator with a self-consistent frequency
$\Omega_{\rm sc}^2 = \Omega_0^2 - 3 \lambda 
\langle x^2\rangle$.

In order to analyze a system coupled to an environment, one should 
generalize the variational principle to the density matrix. Although 
such an extension does exist (see Ref. \cite{jackiw}), we follow 
here an equivalent and simpler method: we replace the renormalized
frequency  by 
the self-consistent one $\Omega_{\rm sc}^2 = {\tilde \Omega}^2 - 3 \lambda 
\langle x^2\rangle$ in the master equation. 
Thus, assuming that the density matrix has the same 
form given in Eq.(\ref{rhogauss2}),
the evolution equations for the real functions
$a(t)$, $b(t)$, and $C(t)$ can be easily obtained
from Eq.(\ref{coopereqs}). The
variational equations then read
\begin{eqnarray}\dot{a} &=& 4 a b - 2 \gamma_0 a + 2 \gamma_0 {\tilde T}
 + \gamma_0 C
\nonumber \\
\dot{b} &=& - 2 a^2 + 2 b^2 - 2 \gamma_0 b + {1\over{2}} C^2 - {1\over{2}} 
{\tilde \Omega}^2 + {3\over{4}} {\lambda\over{(2 a + C)}} \nonumber \\
\dot{C} &=& 4 \gamma_0 a + 4 C b - 2 \gamma_0 C - 4 \gamma_0 {\tilde T}
 \nonumber \\
\dot{N} &=& 2 N b. \label{coopereqs2}\end{eqnarray} 

As before, we are interested in the time dependence of the function 
$2 a - C$.
We have solved numerically Eqs.(\ref{coopereqs2}) for different
values of the parameters. In Fig. 3 we show the time dependence of
$2a - C$ without environment ($\gamma_0=0$, which implies $C=0$). As the
result is an oscillating function, the width of the Wigner function 
and the non-diagonal part of the density matrix do oscillate. The 
self-interacting part of the potential forbids the squeezing of the 
initial state. Therefore, there are no correlations nor decoherence. 
When the coupling to the environment is turned on, $2a-C$ tends to a 
constant of order one. The environment produces the quantum to classical 
transition (Fig. 4).

As has been noted by Cooper et al,
the Gaussian approximation gives  good results 
up to the time where the non-linearities of the potential 
can be neglected. 
To go beyond this point it is necessary to improve 
the approximation.
Cheetham and
Copeland \cite{chetcope} had proposed an improvement based on the following
trial wave function
\begin{equation}\psi (x,t) = {\cal N}(t) e^{-ib x^2} \left[u_0(x,t) + 
a_2 u_2(x,t)
\right],\label{chetham}\end{equation}
where 
\begin{eqnarray} u_0(x,t) &=& \left[{2 a\over{\pi}}\right]^{{1\over{4}}}
 e^{- a x^2} \nonumber \\
u_2(x,t) &=& \left[{a\over{32 \pi}}\right]^{{1\over{4}}}
e^{- a x^2} \left[8 a^2 x^2 - 2\right],\end{eqnarray}
where $a$ and  $b$  are real time-dependent coefficients;
$a_2(t)$ is a complex function. The variational equations read
\begin{eqnarray}\dot{a} &=& 4 a b + \lambda \sqrt{2} {\sin\theta\over{16 a R}}
\nonumber \\
\dot{b} &=& -2 a^2 + 2 b^2 - 1 + 
{7\lambda\over
{8 a}} + {\lambda\sqrt{2}\cos\theta\over{16 a R}}\nonumber \\
\dot{R} &=& \lambda \sin\theta 
{(\cos\theta + R^2\cos\theta+2R\sqrt{2}+
2R^3\sqrt{2})\over{16 a^2 R}} \nonumber \\
\dot{\theta} &=& - 4 a  - {\lambda (4R^3\sqrt{2}\cos\theta + 2R^3
\cos^2
\theta - 2\cos^2\theta + 1 - 6R\sqrt{2}\cos\theta -11R^2)\over{32 a^2 R^2}},
\end{eqnarray}
where we have written $a_2(t) = R e^{i\theta}$.

In Eq.(\ref{chetham}) the zeroth order corresponds 
to the Gaussian approximation. Including only the first non-trivial 
term in the expansion 
in Hermite polynomials, Cheetham and Copeland showed an important
improvement in the results \cite{chetcope}.
For this ``post-Gaussian'' form of the wave funcion, we may write the 
density matrix as
\begin{eqnarray} \rho (x,x',t) &&~= {\cal N}^2(t)\sqrt{{2 a\over{\pi}}} 
e^{- 2 a (\Sigma^2 + \Delta^2)} 
e^{- 4 i b \Sigma \Delta}\nonumber \\
&&\times  \left[1 + 2 \sqrt{2}a (a_2 x'^2 + a_2^* x^2) - 2 \sqrt{2} 
Re a_2 + \sqrt{2} \vert a_2\vert^2 \left(2 a  x^2 x'^2 - x^2 - x'^2 
+ {1\over{2 a}}\right)\right]\end{eqnarray} 
which produces a complicated Wigner function
\begin{eqnarray}W(x,p,t) &=& {{\cal N}^2(t)\over{2\pi}}\sqrt{{2a\over{\pi}}}  
e^{-{(p - 2 b x)^2\over{2a}}} e^{-2 a x^2} \left[1 + B(t) + D(t) p + E(t) p^2 
+ F(t) p^4\right.  \nonumber \\
&+&\left.  H(t) x + 
I(t) x p + J(t) x p^3 + K(t) x^2 + L(t) x^2 p^2 + M(t) x^3 p + P(t) x^4\right]
\end{eqnarray}
where the capital letters are functions of $a_2$, $a_2^*$, $a$, and $b$.

The Wigner 
function is clearly non positive definite. To illustrate this fact,
we have solved numerically the variational equations for a
Gaussian initial state. In Fig. 5
we show the Wigner function for a time where the wave 
function is no longer Gaussian. The non-Gaussian shape of the
wave function produces a non positive Wigner function.

It is in principle possible to include the environment
in the improved version of the Gaussian approximation. However, we will
not follow here this possibility. Instead, we will include it in the exact 
numerical solution to the problem.

\section{Including self-interaction: exact numerical solution}

In order to get a complete answer about the 
quantum to classical transition for a double well potential,
it is necessary to solve exactly the 
master equation given in Eq.(\ref{master}) (adding the $\lambda x^4$ term 
to the Hamiltonian $H_{\rm syst}$ in the first term of the rhs). 
The Fokker-Planck equation (\ref{fokker}) for the reduced Wigner function 
has an additional term coming from the non-linearities of the potential,

\begin{eqnarray}\dot{W_r}(x,p,t) =&& -\{H_{\rm syst}- {1\over{2}} M 
{\tilde\Omega}^2(t),W_r\}_{\rm PB} +
 2 \gamma (t) \partial_p(pW_r)
+ D(t) \partial^2_{pp}W_r \nonumber \\
&-& f(t) \partial^2_{px}W_r - {\lambda\over{4}} x 
\partial^3_{ppp}W_r,
\label{fokker2}\end{eqnarray}
 
We have solved numerically  this Fokker-Planck equation, in the high 
temperature limit, for different values
of the diffusion coefficient $D$, in order to illustrate its relevance in the
quantum to classical transition. We have chosen
as initial condition a Gaussian state centered at $x_0=p_0=0$ 
with minimal uncertainty ($\sigma_x^2 = 0.5$ and $\sigma_p^2 = 
0.5$). The Wigner function is initially positive definite, 
and  different from zero only
near the top of the potential.
We have set the 
coupling constant $\lambda = 0.01$, the renormalized frequencies ${\tilde 
\Omega} = 
\omega_n = 1$ (we are measuring time in units of ${\tilde \Omega}$) and the 
bare masses also equal to one.
 
It is illustrative to examine first the exact result when the
environment is absent (for this case we have solved numerically 
the Schr\"odinger equation). The initially Gaussian Wigner function
begins to squeeze in the $x=p$ direction and, before 
the spinodal time ($t_{\rm sp}\sim 2.3$) it 
becomes a non-positive function (Fig. 6). During the evolution, the 
Wigner function covers all the phase space  (Fig. 7) and it is  clear 
that it is not possible to consider it as a classical probability 
distribution. Although we started with 
a special initial state (Gaussian  with 
minimum uncertainty), the non-linearities of the potential make
the Wigner function a non-positive distribution.

Let us now consider a coupling with an environment such that
the normal 
difusion coefficient is $D = 0.01$. As expected, the evolution
of the Wigner function is similar to the previous one at early times (Figs. 
8 and 9). However, as can be seen from Figs. 10 and 11, at long times 
it becomes positive definite and peaked around the classical phase space. 
  
The effect of the environment is more dramatic for larger values of 
the diffusion coefficient (see Figs. 12 - 15). For $D=0.1$, the 
Wigner function
is almost positive definite for $t\sim 2 t_{\rm sp}$ (Fig. 13). 
In our last example, $D=1$, the quantum to classical transition 
takes place almost instantaneously, even before
the quantum particle pass through the spinodal point (Figs. 15 and 16).

It is 
interesting to note that, as the difusion coefficient grows, the 
amplitude of the Wigner function falls down. This is due to the 
fact that the decoherence increases with $D$. The reduced density
matrix diagonalizes. As a consequence, its ``Fourier transform",
the reduced Wigner function, spreads out.
 
Our numerical results show explicitly
that the existence of the environment is 
crucial in the quantum to classical transition. The decoherence time
depends on the temperature and the coupling between system and environment
through the difussion coefficient $D$.
 
Both the Schr\"odinger equation for the closed problem and the Fokker-Planck 
equation (\ref{fokker}) were numerically solved using a fourth-order spectral 
algorithm \cite{feit}. Numerical checks included carrying out simulations at 
different spatial and temporal resolutions.

\section{Phase transitions in field theory}

In any field theory where the classical potential has a local maximum at
$\phi =0$, the long wavelength modes of the field are unstable, and grow 
in time. 
From these modes it should be possible to identify
the classical field that plays the role of order parameter of the 
transition to the minimum of the potential. 
The emergence of semiclassical coherent, large 
amplitude field configurations should be a consequence 
of time evolution. 
Therefore, we need a quantum field description 
of the dynamics, for the early stages of phase ordering and growing 
of long-wavelength fluctuations, as well as for the classicalization  and 
decoherence of such fluctuations. 

There have been different approaches to this problem in the literature. Many
works assume that the field can be split as $\phi (x) = \phi_0(t) +
\hat\phi (x)$, where $\phi_0$ is the mean value of the field and  
$\hat\phi$ are
the quantum fluctuations \cite{mean}. This can be only an approximation 
to the full 
problem, since by symmetry arguments it is obvious that the mean value of
the field must vanish. One should think of $\phi_0(t)$ as the mean value
of the field inside one of the domains where the phase transition is taken 
place.
In a more realistic approach, the field is split as \cite{nos} $\phi (x) = 
\phi_<(x) +
\phi_> (x)$, where $\phi_<$ and $\phi_>$ describe the short and 
long-wavelength
modes of the field. Hopefully, the effective dynamics of the long-wavelength
modes should indicate a quantum to classical transition. 
A third possibility is
to address this problem by analyzing the quantum dynamics of the 
full field, assuming a vanishing mean value, and check that some of the 
modes become classical variables during the dynamical evolution.

The main technical complication comes from the fact that,
as has  been pointed out, the initial  growth of the quantum
fluctuations is so important that a non-perturbative treatment 
is unavoidable \cite{boya1}.
For this reason, people have considered the so called Gaussian 
approximation or,
alternatively, the large $N$ limit of $O(N)$ models with spontaneous 
symmetry 
breaking. In both approximations one assumes that the wave function 
associated
to the different modes of the quantum field is a 
Gaussian function, with  a self-consistent
set of parameters.

For concretness, let us describe the work of
Boyanovsky et al \cite{boya2}. Consider an $O(N)$ field theory
\begin{equation}\vec \Phi(\vec x, t) = \left(\vec\Phi_1(\vec x, t), 
\vec\Phi_2(\vec x,t), ....., \vec\Phi_N(\vec x, t)\right),\end{equation}  
with a potential given by
\begin{equation}V[\vec\Phi ] = {1\over{2}} m^2(t) \vec\Phi . \vec\Phi
+ {\lambda\over{8N}} [\vec\Phi . \vec\Phi ]^2,\end{equation}
where $m^2(t)$ becomes negative for $t>0$. We will focus on 
the case in which 
the initial state is symmetric, i.e. $\langle\Phi\rangle = 0$.

The Hamiltonian is 
\begin{equation}H = \sum_{\vec k} \left\{{1\over{2}} \vec\Pi_{\vec k}.
\vec\Pi_{- \vec k} + {1\over{2}} W^2_{k}(t) \vec\Phi_{\vec k} .
 \vec\Phi_{-\vec k}\right\},\end{equation}
where $\vec\Phi_{\vec k}$ is the spatial Fourier transform of the 
field, and
 $W_k$ is defined as
\begin{equation}W_k^2 = m^2(t) + k^2 + {\lambda\over{2N}} \int 
{d^3k\over{(2\pi)^3}} \langle \vec\Phi_{\vec k} . \vec\Phi_{-\vec k}
\rangle (t).\end{equation} 
We start with 
a Gaussian initial state, 
and we assume that the wave functional will be always Gaussian, 
describing 
a pure quantum mechanical state, 
\begin{equation}\Psi[\vec\Phi ,t]= \prod_{k}\left\{N_k(t) e^{-{A_k(t)
\over{2}}\vec\Phi_{\vec k} . \vec\Phi_{-\vec k}}\right\},
\label{gaufunc}
\end{equation}
where $A_k(t=0) = W_k(t<0)$. 

The functional Schr\"odinger equation gives the following differential 
equations for $N_k$ and 
$A_k$, 
\begin{eqnarray}{d\over{dt}}\ln N_k(t) &=& -{i\over{2}}A_k(t)\nonumber \\
i{dA_k(t)\over{dt}} &=& A^2_k(t) - W_k^2(t).\end{eqnarray}
Introducing the notation
\begin{equation}A_k(t) = -i {\dot{\phi}_k(t)\over{\phi_k(t)}},
\end{equation}
the dynamical equation in the $1/N$ approximation becomes 
\begin{equation}\ddot{\phi}_k(t) + W^2_k(t) \phi_k(t) = 0.\end{equation}
The expectation value of $\vec\Phi_k(t)^2$ in the state of Eq.(\ref{gaufunc}) 
is given by
\begin{equation}\langle\vec\Phi_{\vec k}.\vec\Phi_{-\vec k}\rangle (t) = N 
\vert \phi_k(t)\vert^2.\end{equation}
Therefore we have a self-consistent system for $t>0$ given by 
\begin{eqnarray}&&\ddot{\phi}_k(t) + \left[k^2 + M^2(t)\right]\phi_k(t) = 0
\nonumber \\
&&M^2(t) = - \mu^2 + {\lambda\over{2}} \int {d^3k\over{(2\pi)^3}} \vert 
\phi_k(t)\vert^2.
\label{modeq}
\end{eqnarray} 

The numerical solution of these equations reveals the following picture:
at early times, the long-wavelength fluctuations ``see'' an inverted 
oscillator potential, and grow exponentially in time. 
This is a linear regime where the self-interaction can be
neglected. At intermediate times
the backreaction of these fluctuations is as important as the classical
terms in the Lagrangian. This period is highly non-linear. At long times
times the effective mass of the fluctuations vanishes asymptotically 
\cite{boya3}.

It can be seen from the wave functional Eq.(\ref{gaufunc}) 
that, asymptotically, 
the long-wavelength modes become classically correlated. This is analogous
to the situation described by the inverted harmonic oscillator. When
$M^2\sim 0$, the mode equations (\ref{modeq}) can be easily solved 
\cite{boya4}
\begin{equation}
\phi_k = a \cos kt + b {\sin kt\over{k}}.
\end{equation}
As a consequence, the width of the Gaussian wave function 
for the mode $k$
increases linearly with time $Re{A_k}^{-1}\sim t$, when $ k t << 1$. 
As we have seen, a Gaussian wave function
has a positive definite Wigner function associated to it. Moreover, 
as the width of the wave function (or the density matrix) increases, 
the Wigner function becomes sharply peaked around the classical
trajectory. This is indeed what happens for $k<<1/t $. These modes become
classically correlated. 

In view of the results in the previous sections of this paper, we see that
in the large $N$ limit, and at long times, the  dynamical evolution
of the $O(N)$ model shows ``classical correlations'' but not a full
quantum to classical transition. As for the inverted oscillator
without environment, 
the density matrix does not become
diagonal. The ``correlation time'' depends only on the details
of the classical potential. 
{\footnote{There is a quantitative difference with
the example we presented in Section II: there
the width
of the wave function increased exponentially. Here, as the mode
become massless, the growth of the width is only linear in time
(an inverted oscillator would correspond to a negative mass).}}

On the other hand, these classical correlations
depend crucially on the Gaussian form of the wave function. To illustrate
this point, we consider  another example from non relativistic
quantum mechanics, the d-dimensional quantum roll, which
corresponds to the zero space dimensional limit of the $O(d)$ field theory.

The model is described by the Hamiltonian \cite{mihaila}
\begin{equation} H = -{1\over{2}} \nabla^2 + V(r),\end{equation}
where
\begin{eqnarray}\nabla^2 &=& \sum_{i=1}^{N}{\partial^2\over
{\partial x_i^2}} ~~
, ~~~~ r^2 = \sum_{i=1}^{N}x_i^2 \nonumber \\
V(r) &=& {\lambda\over{8d}} (r^2 - r_0^2)^2.\end{eqnarray}
The Schr\"odinger equation reads
\begin{equation}H \Psi(x_i,t) = i {\partial\Psi(x_i,t)\over{\partial t}}.
\end{equation}
This problem can be studied in a multidimensional coordinate system with 
$r$ the radial coordinate and a set of $d-1$ 
angular coordinates, such that the Laplacian is
\begin{equation}\nabla^2 = {\partial^2\over{\partial r^2}} + {(d-1)\over{r}} 
{\partial\over{\partial r}} - {L^2_{d-1}\over{r^2}}.\end{equation}
Here $L^2_{d-1}$ is the generalized orbital angular momentum 
operator. Starting with a Gaussian, radially symmetric initial state centered
at the top of the hill ($r=0$), 
there angular momentum will vanish. Therefore 
the Schr\"odinger equation can be re-written as
\begin{equation} {\tilde H}(r,l) \Phi (r,t) = i {\partial \Phi (r,t)
\over{\partial t}},\end{equation}
where 
\begin{eqnarray}\Phi (r,t) &=& r^{{(d-1)\over{2}}} \Psi (r,t)\nonumber \\
{\tilde H} &=& - {1\over{2}} {\partial^2\over{\partial r^2}} + U(r)
\nonumber \\
U(r) &=& {(d-1) (d-3)\over{8 r^2}} + {\lambda\over{8d}} (r^2 - r_0^2)^2.
\end{eqnarray}    
Following Mihiala et al \cite{mihaila}, we have solved the Schr\"odinger 
equation 
numerically. We took an initial Gaussian state centered at the potential 
top, and used $d=20$, $r_0 = 6.4$, and $\lambda = 10$. In Fig. 17 we 
plot $<r^2/d>$. After a period of oscillations, this function reachs 
a constant value for large times. 
This asymptotic value would imply a vanishing 
effective mass in the large $d$
approximation. However, as can be seen from Fig. 18, the Wigner function
is not positive definite even when the effective mass vanishes.
The reason for this is that the wave function that describes
the d-dimensional slow roll has a complicated structure, and cannot
be approximated by a Gaussian function (see Fig. 19).

The lesson we learned from this example is that, in order to get a
Wigner function that is positive definite and peaked around a classical
trajectory at long times, it is necessary to have both vanishing 
effective mass and
a Gaussian wave function.
Therefore, it is quite possible that in a
field theory calculation with finite $N$ there will be no classical
limit nor classical correlations unless one considers the coupling of
the field with an environment. The presence of the environment will
introduce a new temporal scale, the decoherence time, which will
indicate when the order parameter becomes a classical variable. For 
each mode, the decoherence time could be shorter than the spinodal 
time, allowing for a classical description of these modes in the 
non-linear regime. 
A realistic treatment of the environment seems to be crucial to
understand the classical limit in field theory phase transitions. 

\acknowledgments
This work was supported by Universidad de Buenos Aires, Conicet (Argentina)
and Fundacion Antorchas.

\newpage
{\bf Figure Captions}

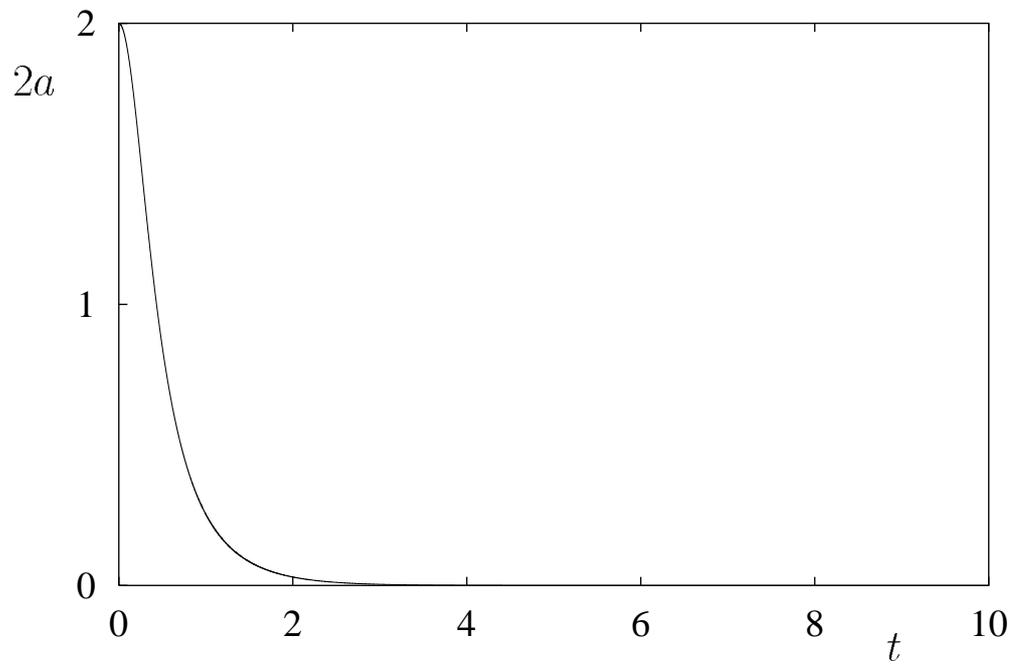
{\bf Figure 1}: Function $2 a$ which gives the width of the Wigner function 
for the upside down harmonic potential. 
 
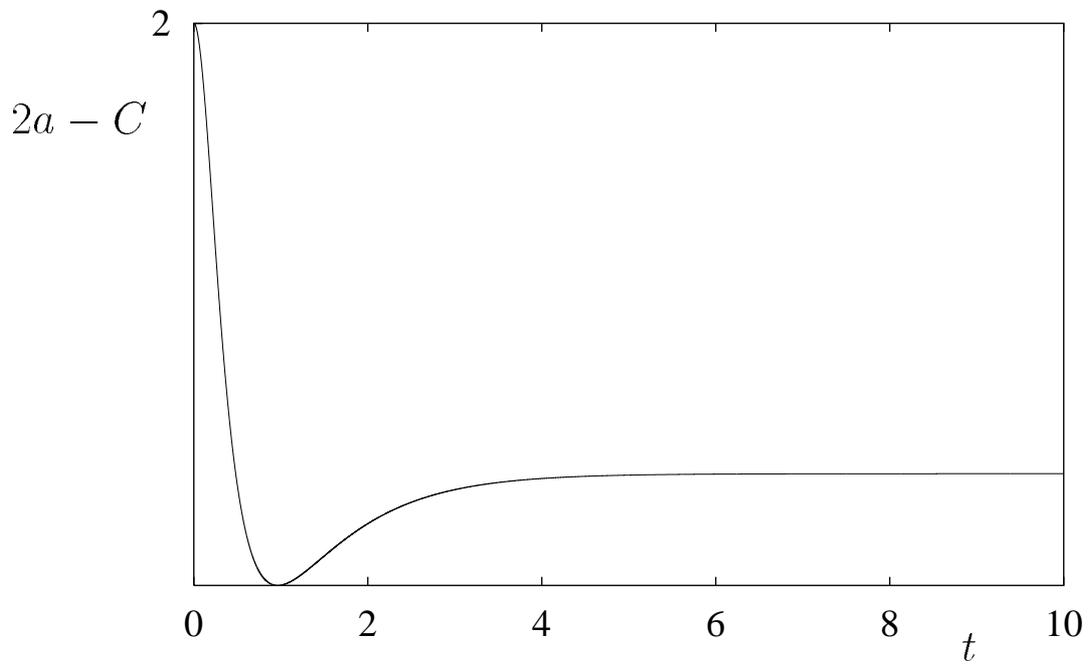
{\bf Figure 2}: Width of the Wigner function when an environment is 
taking into account. We considered an underdamped case with $\gamma_0 = 0.001$
and ${\tilde T} = 500$.

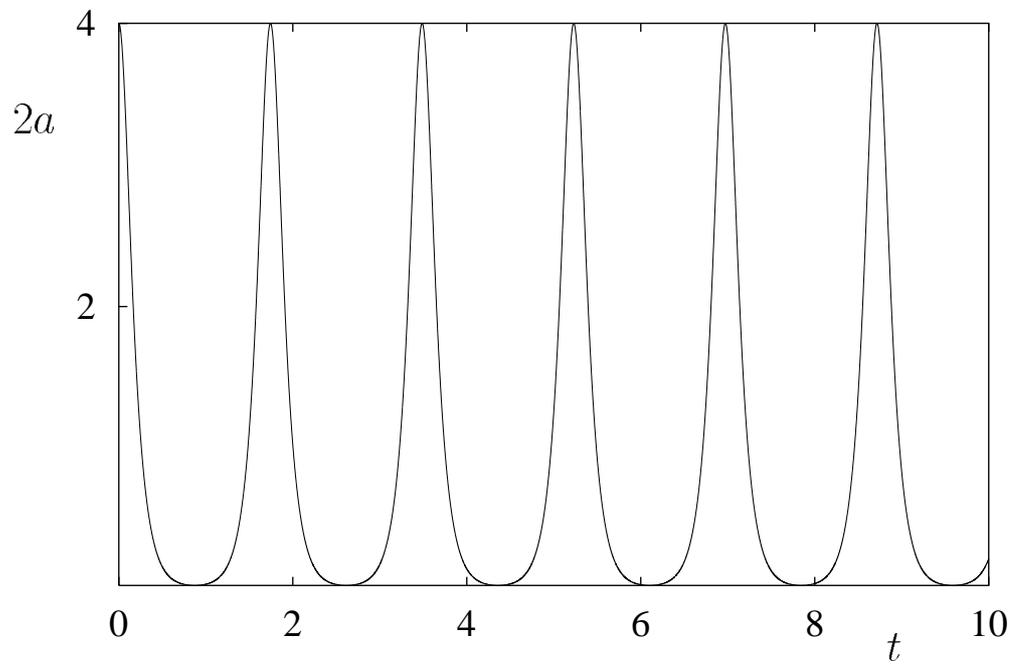
{\bf Figure 3}: Time dependent width of the Wigner function for a 
self-interacting ($\lambda x^4$) potential, using the Gaussian approximation.

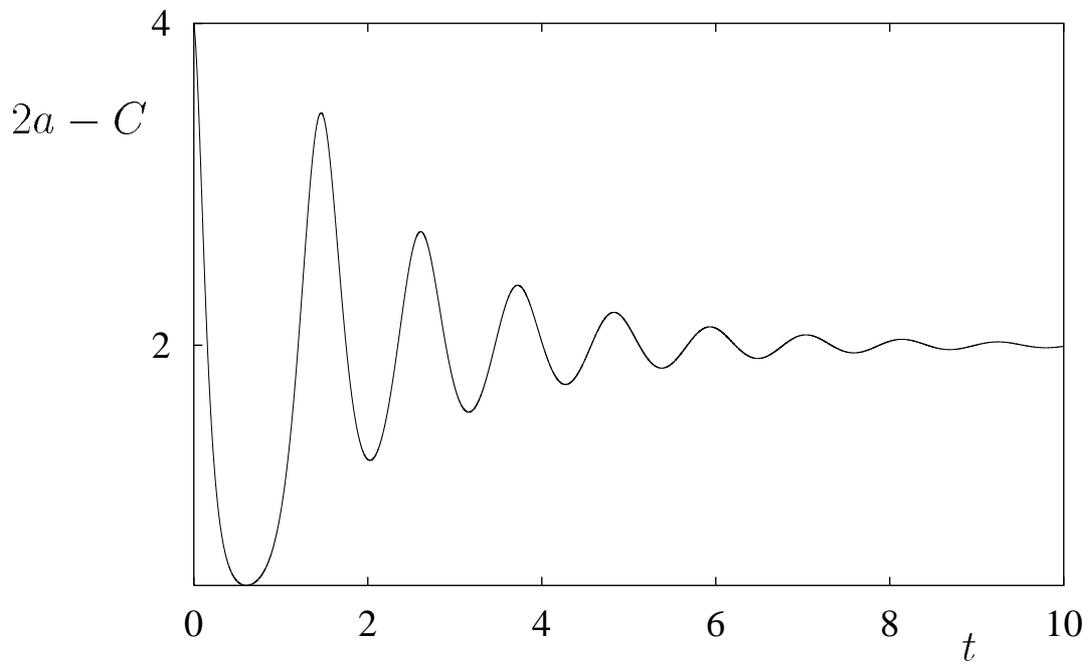
{\bf Figure 4}: Same as figure 3 but including an underdamped ($\gamma_0 =
 0.001$) environment with ${\tilde T} = 50$.

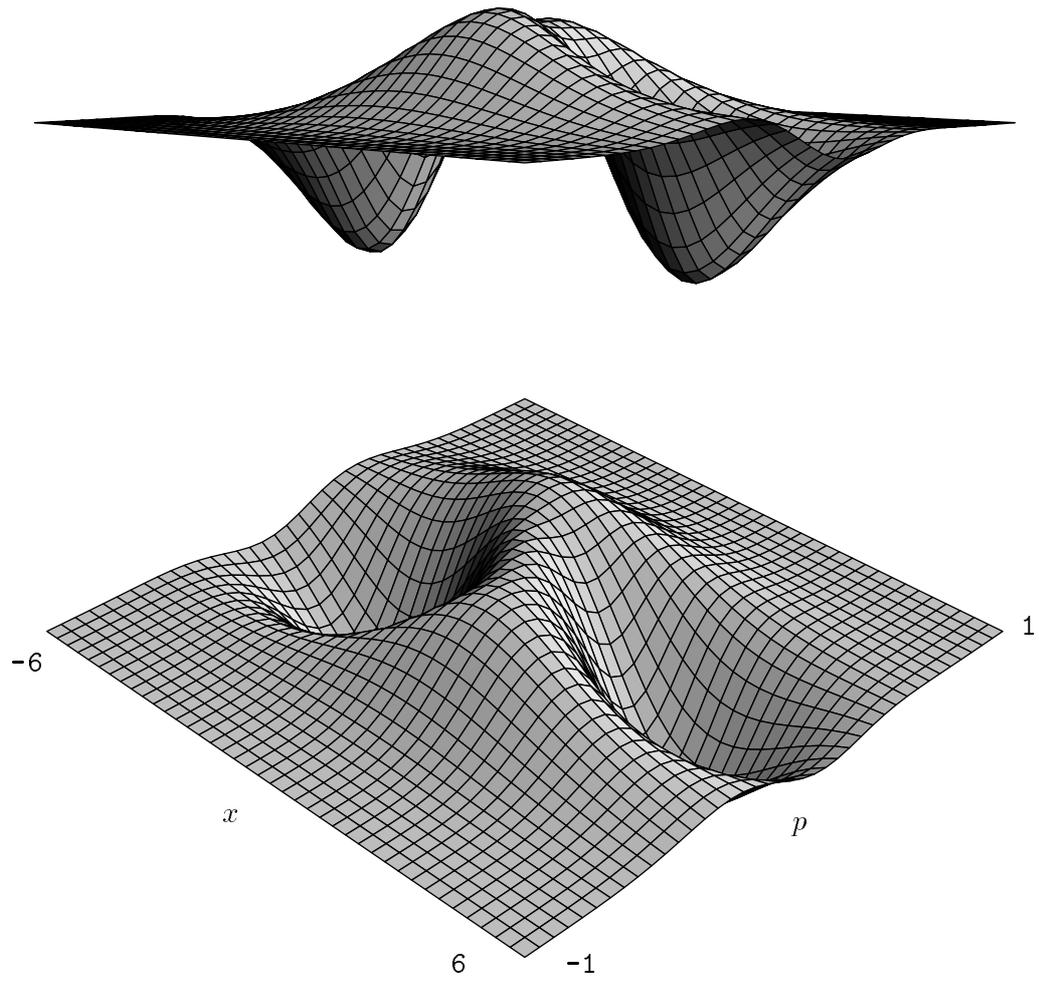
{\bf Figure 5}: Wigner function for the post-Gaussian approximation. It is 
evaluated at a time ($t = 20$) when the wave function is no longer a Gaussian.

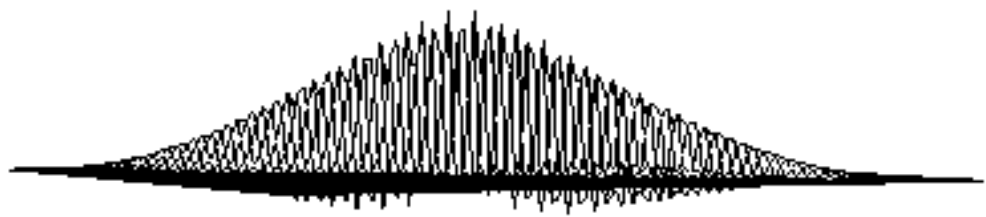
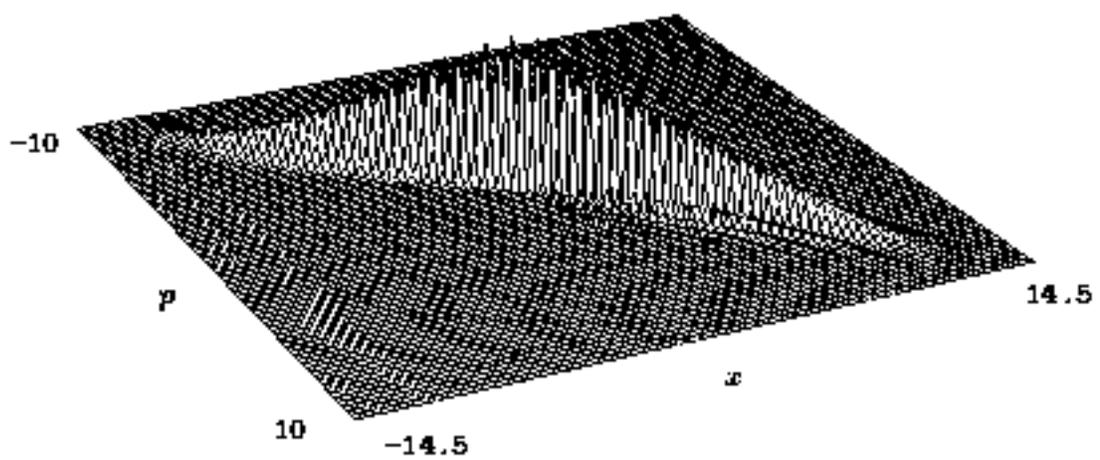
{\bf Figure 6}: Wigner function for $t = 2 < t_{\rm sp} = 2.3$, no 
environment is considered.  

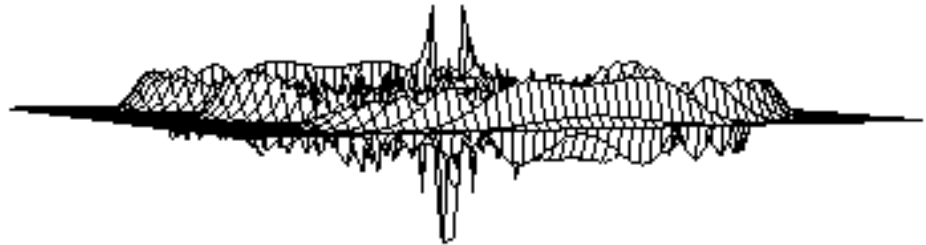
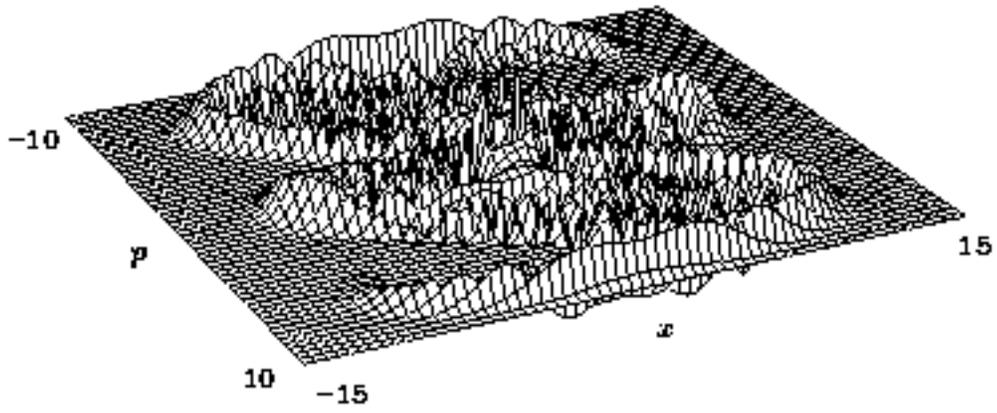
{\bf Figure 7}: Wigner function for $t=4$. As the function is not 
positive definite, there is no correlation.

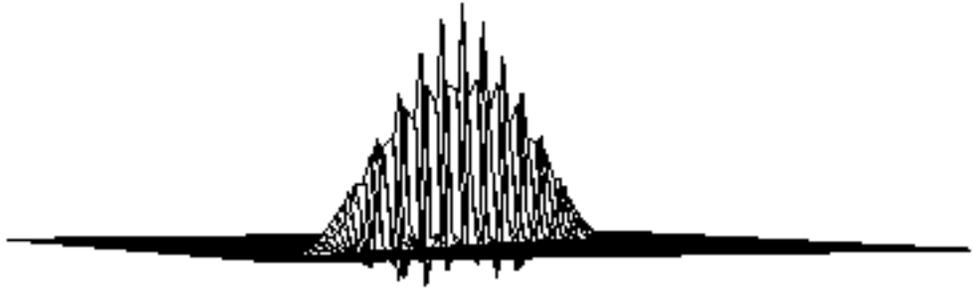
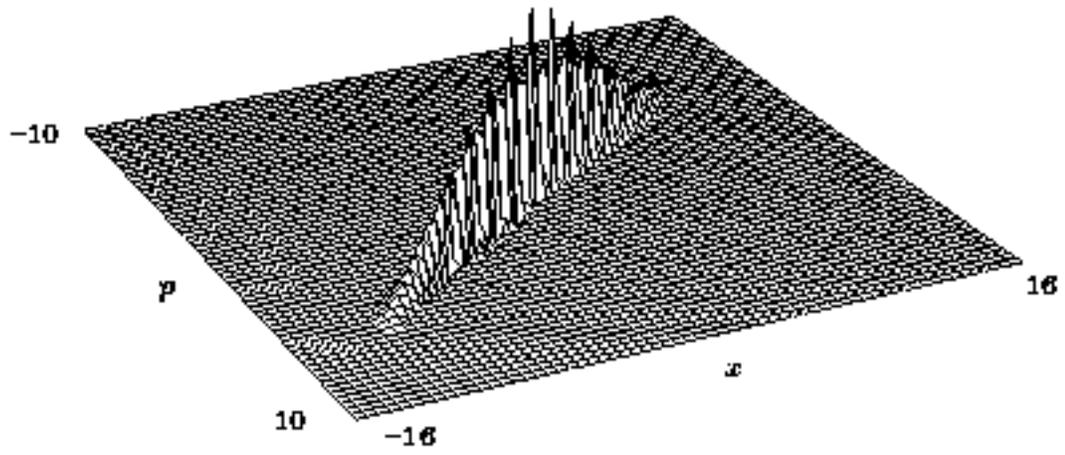
{\bf Figure 8}: Wigner function for $t=2$ including an environment 
with diffusion coefficient $D= 0.01$. 

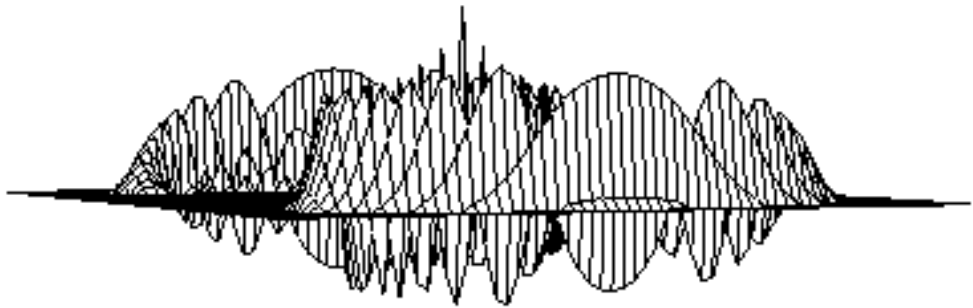
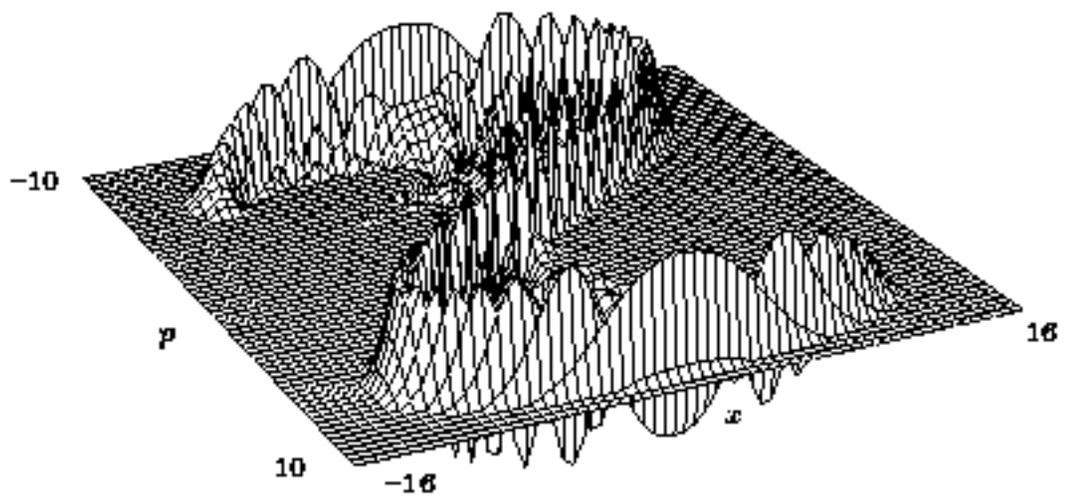
{\bf Figure 9}: Same Wigner function for time $t = 4$. The
Wigner function is not positive definite.

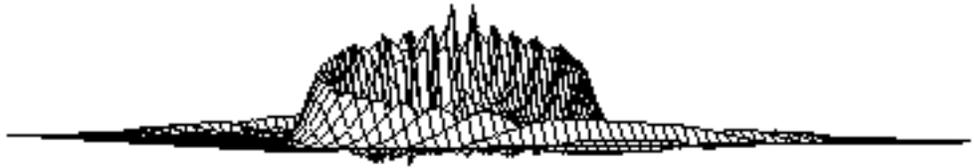
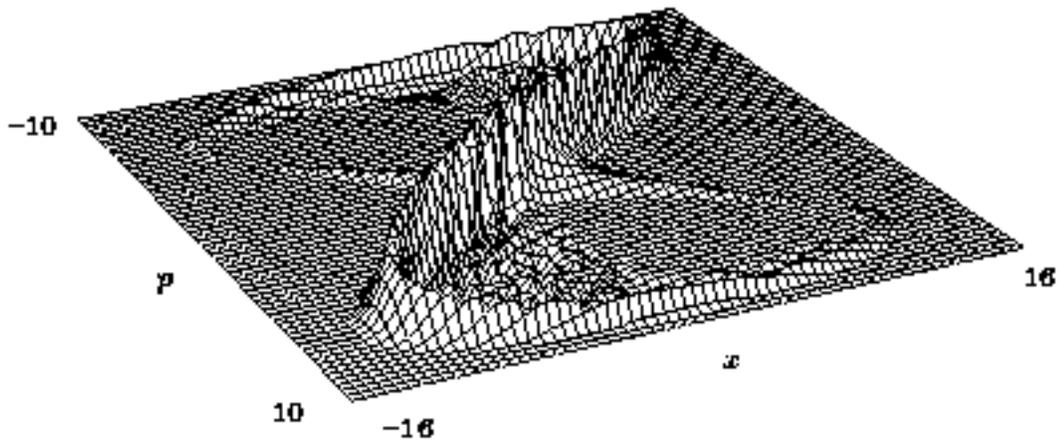
{\bf Figure 10}: Same Wigner function for $t=10$.

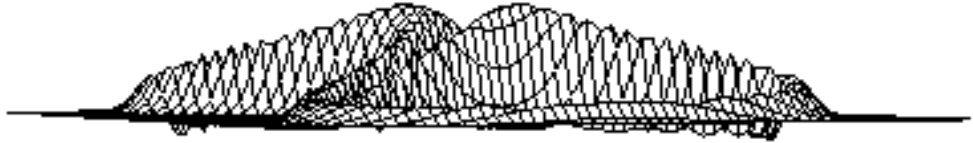
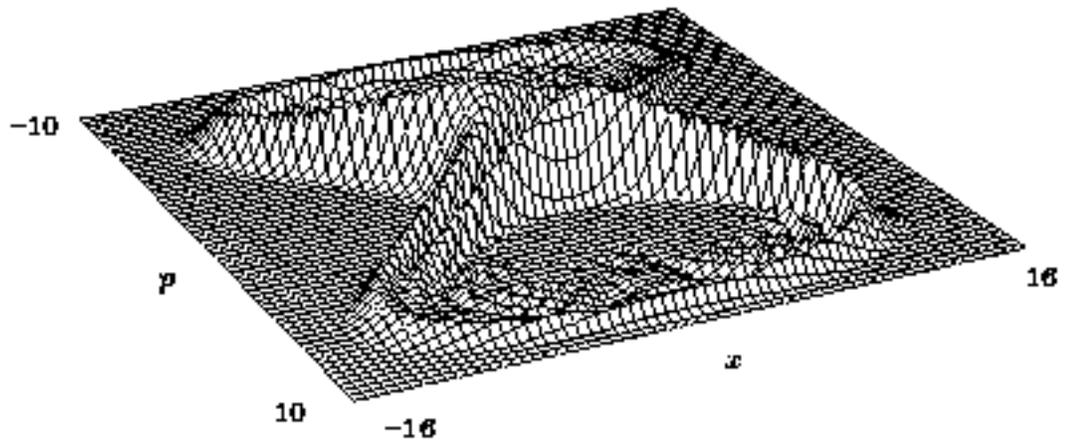
{\bf Figure 11}: Same Wigner function for $t = 15$. Only for $t >> t_{\rm sp}
$ we have classical correlations and a positive Wigner function.

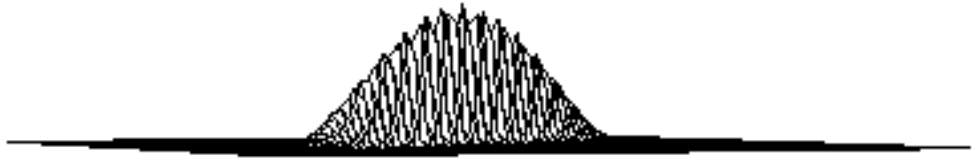
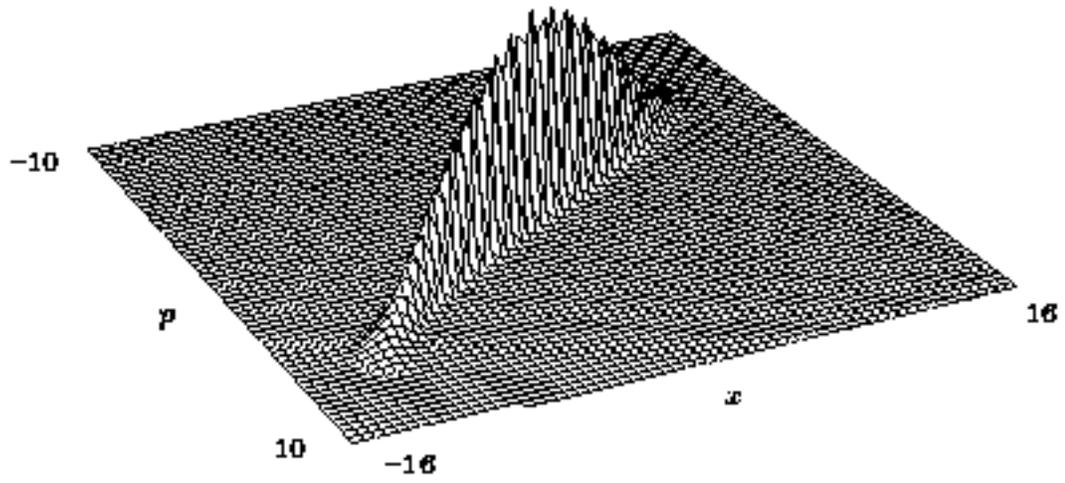
{\bf Figure 12}: Early time ($t = 2$) Wigner function for the case $D=0.1$.

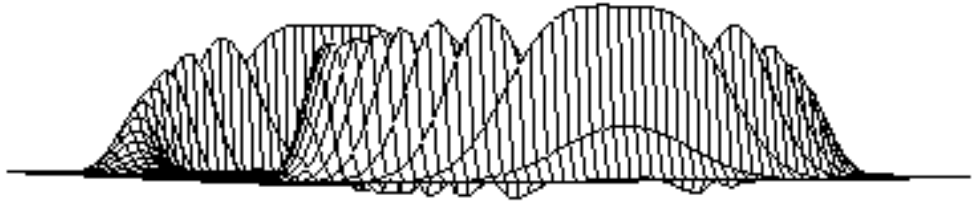
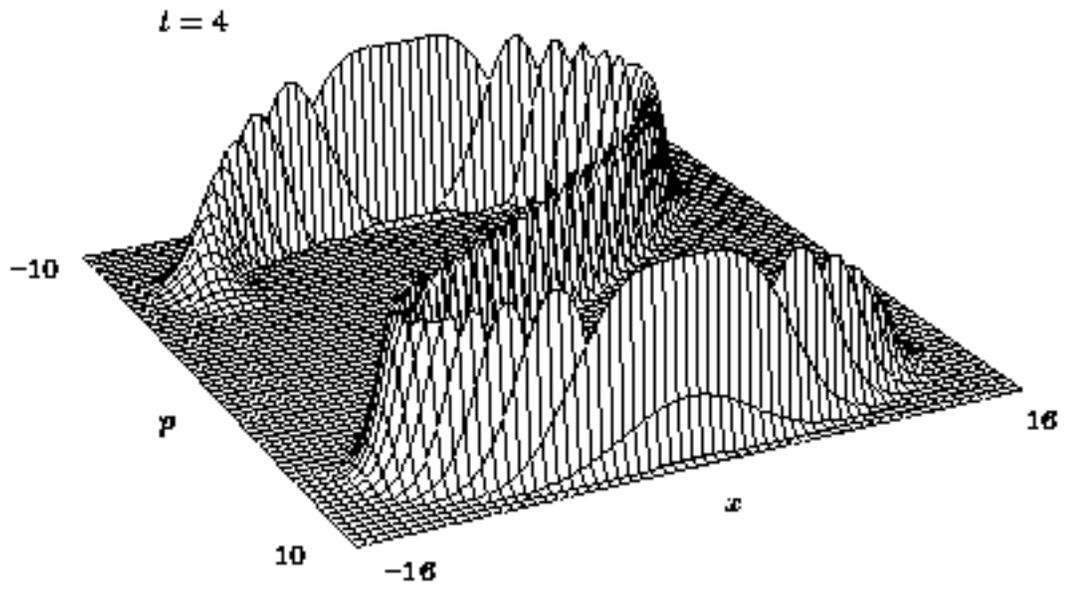
{\bf Figure 13}: Same as figure 12, for $t = 4$.

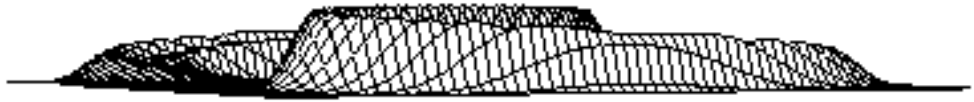
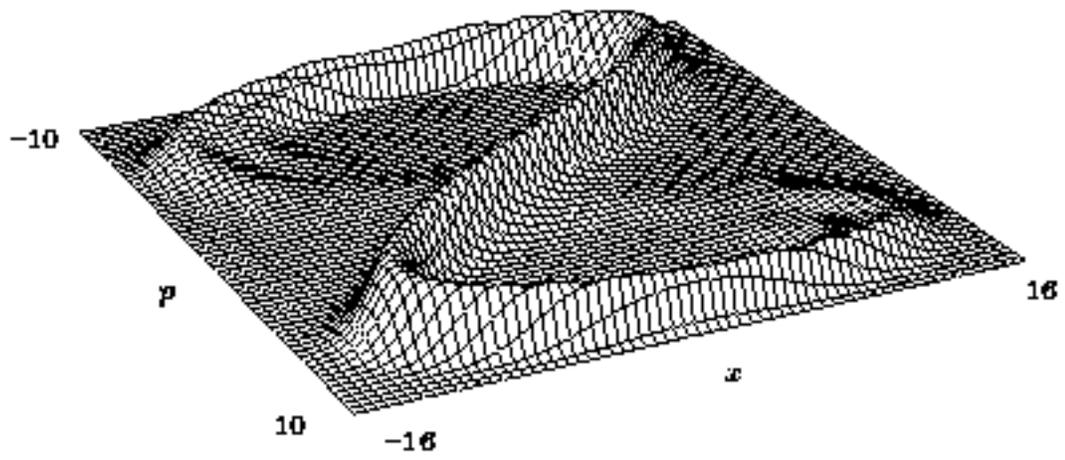
{\bf Figure 14}: Long time ($t= 10$) behaviour of the Wigner function 
for $D=0.1$. Here we can consider the system as classical.

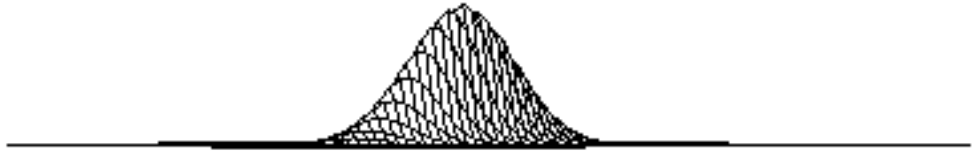
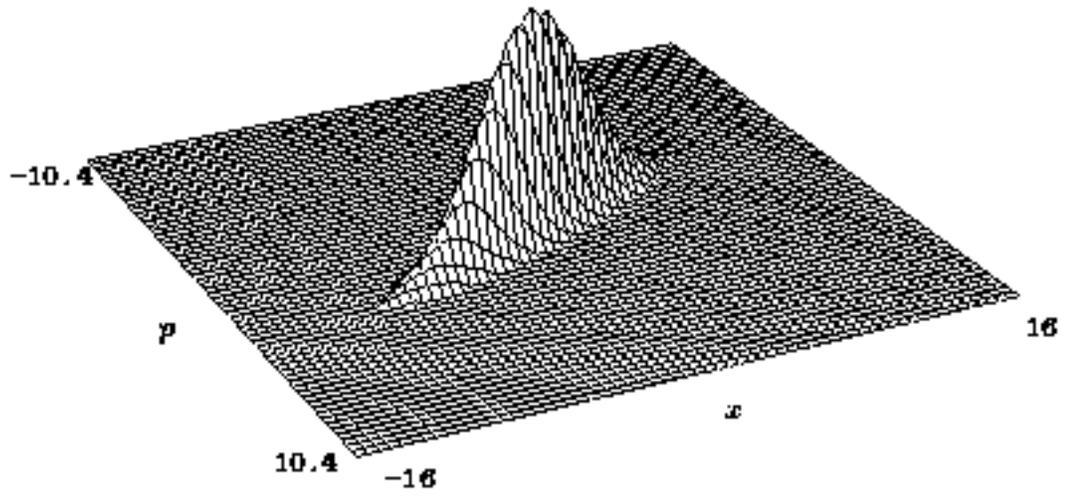
{\bf Figure 15}: Early time ($t = 1$) Wigner function for a  diffusion 
coefficient $D=1$.

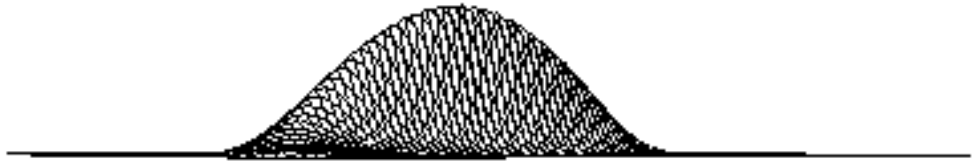
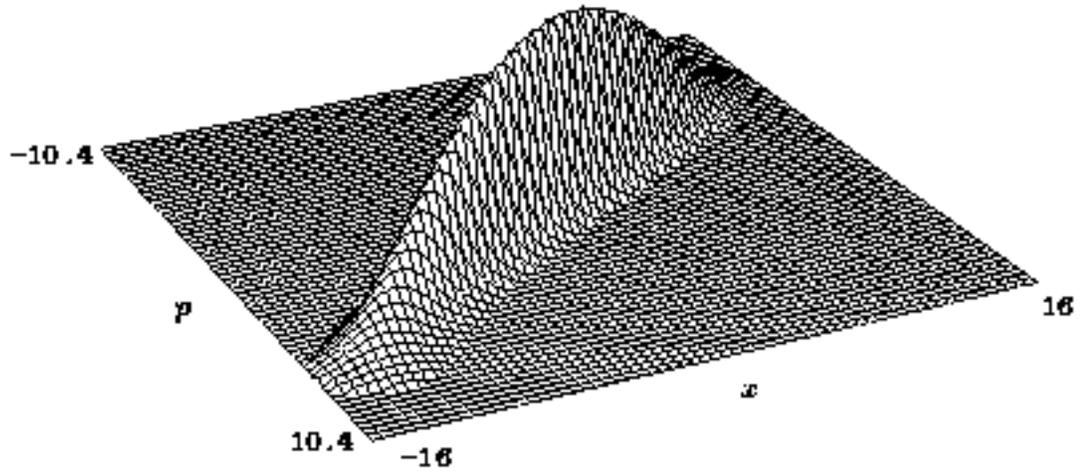
{\bf Figure 16}: Same as figure 15 for $t = 2$. Classicality emerges 
before the spinodal time. 

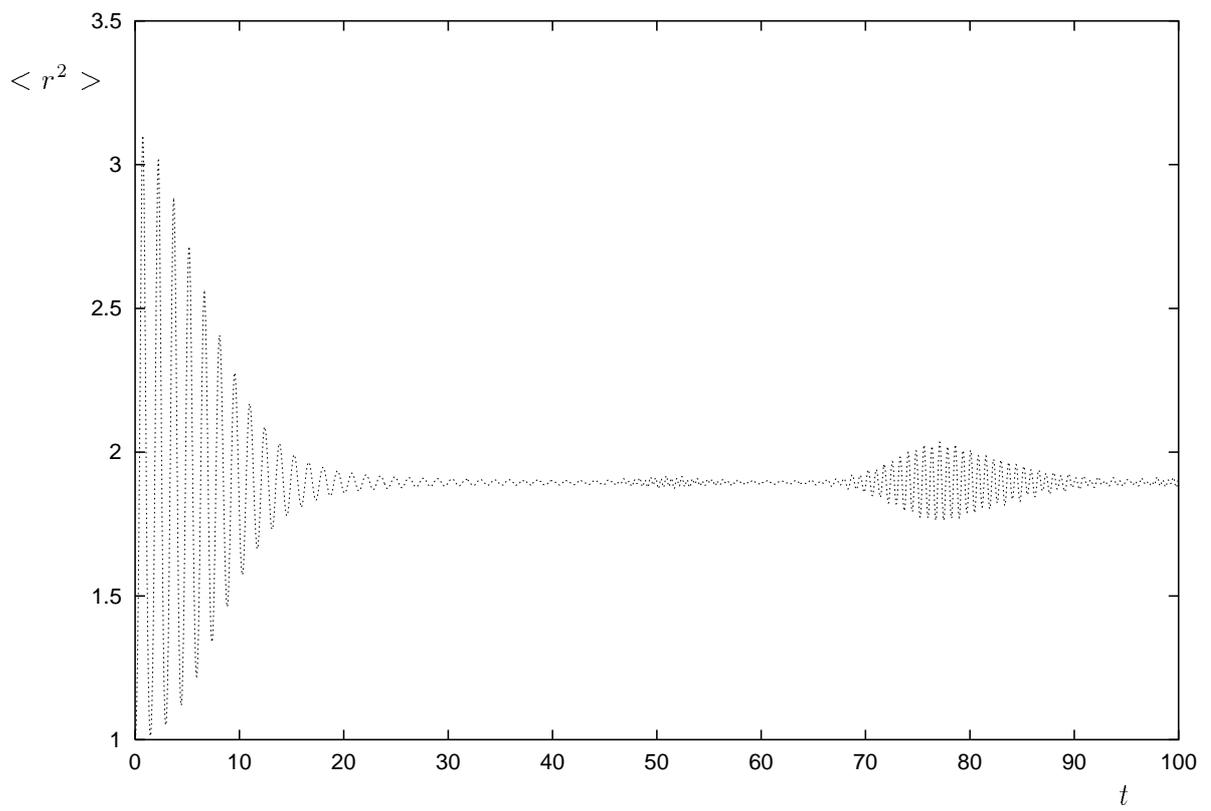
{\bf Figure 17}: d-dimensional quantum roll. $\langle r^2/d\rangle$ 
for $d = 20$, $r_0=6.4$, and $\lambda = 10$.  

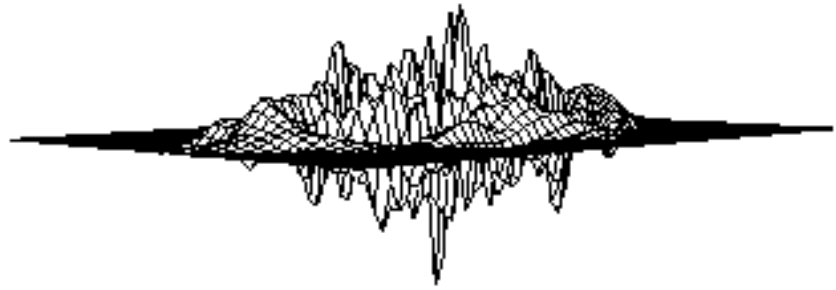
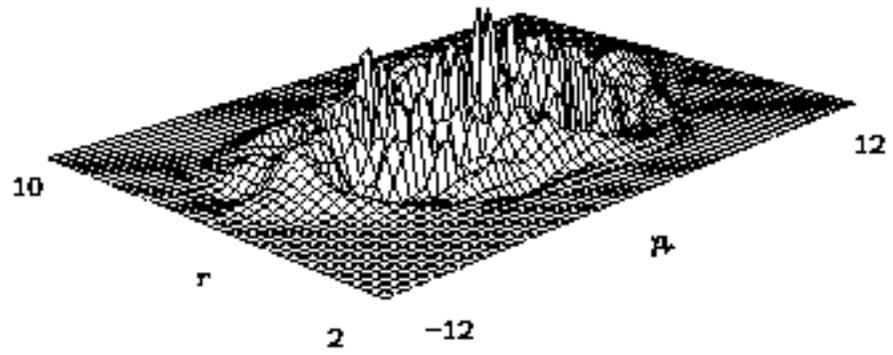
{\bf Figure 18}: Wigner function for the d-dimensional 
example, evaluated at a time ($t = 40$) when the ``effective mass'' is zero.

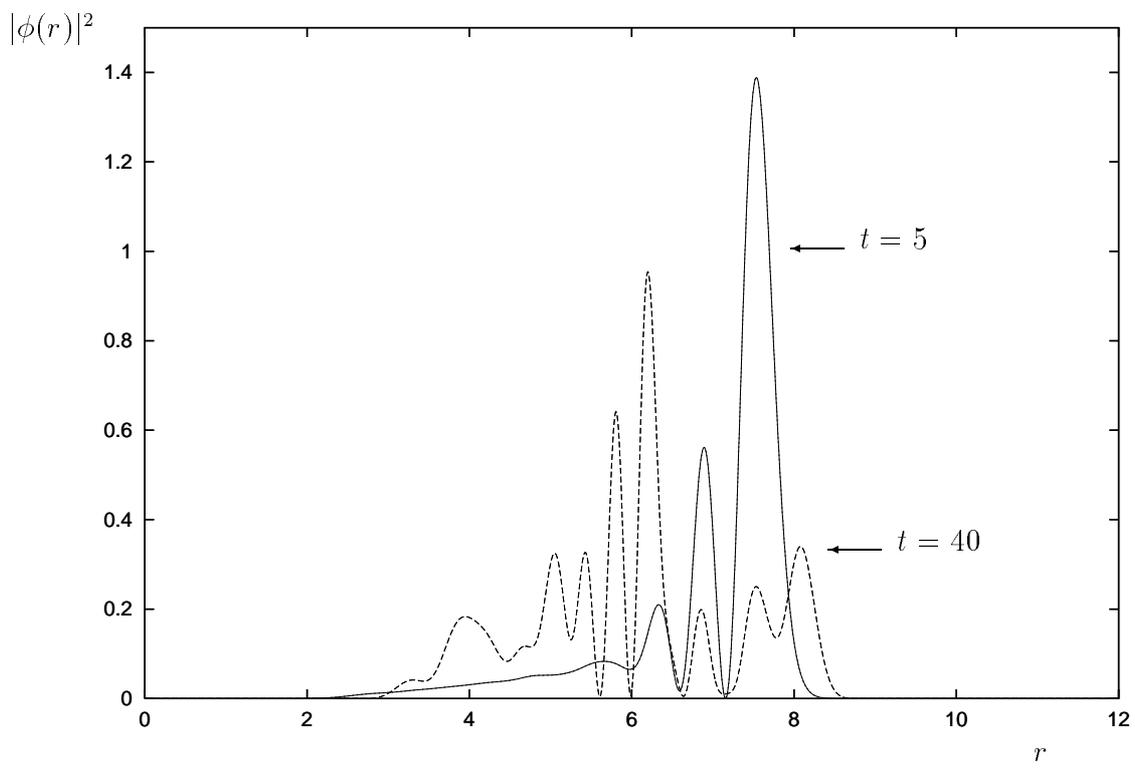
{\bf Figure 19}: $\vert \phi (r, t)\vert^2$ for the d-dimensional example. 
The wave function is clearly non Gaussian.


\begin{references} 

\bibitem{giulmol} See for example, {\it Decoherence and the appearance of
a classical world in quantum theory}, D. Giulini et al, Springer
Verlag (1996); K. Molmer, Phys. Rev. {\bf A55}, 3195 (1997); J. Mod. Optics 
{\bf 44}, 1937 (1997) 

\bibitem{unruhzu} W.G. Unruh and W.H. Zurek, Phys. Rev. {\bf D40}, 1071 (1989)
\bibitem{zurek} W.H. Zurek, Nature {\bf 317}, 505 (1985); Phys. 
Rep. {\bf 4}, 
276 (1996) and references therein

\bibitem{rivers} G. Karra and R.J. Rivers, Phys. Lett. {\bf B414}, 28 (1997); 
R.J. Rivers, 3rd Colloque Cosmologie, Observatoire de Paris, in the 
Proceedings edited by H.J. de Vega and N. S\'anchez, 314, World Scientific 
(1995); A.J. Gill and R.J. Rivers, Phys. Rev. {\bf D51}, 6949 (1995); 
G.J. Cheetham, E.J. Copeland, T.S. Evans, and R.J. Rivers, Phys. Rev. {\bf D
47}, 5316 (1993); G.J Stephens, E.A. Calzetta, B.L. Hu, and S. Ramsey, 
Phys. Rev. {\bf D59}, 045009 (1999) 

\bibitem{nos} F.C. Lombardo and F.D. Mazzitelli, Phys. Rev. {\bf D53}, 2001 
(1996); see also C. Greiner and B. Muller, Phys. Rev. {\bf D55}, 1026 (1997)

\bibitem{hpz} B.L. Hu, J.P. Paz, and Y. Zhang, Phys. Rev. {\bf D45}, 2843 
(1993); {\bf D47}, 1576 (1993); J.P. Paz, S. Habib, and W.H. Zurek, Phys. Rev.
 {\bf D47}, 488 (1993) 


\bibitem{guthpi} A. Guth and S.Y. Pi, Phys. Rev {\bf D}32, 1899 (1985) 

\bibitem{laflamme} R. Laflamme and J. Louko, Phys. Rev. {\bf D43}, 3317 (1991)

\bibitem{pazzurek} W.H. Zurek and J.P. Paz, Phys. Rev. Lett. {\bf 72}, 2508 
(1994)

\bibitem{kiefer} C. Kiefer, J. Lesgourgues, D. Polarski, and A.A. Starobinsky, 
Class. Quant. Grav. {\bf 15}, L67 (1998); C. Kiefer, D. Polarski, and A.A. 
Starobinsky, gr-qc/9910065 and references therein

\bibitem{jackiwk} R. Jackiw and A. Kerman
Phys. Lett. 71{\bf A}, 158 (1979)

\bibitem{cooper} F. Cooper, S.Y. Pi, and P. Stancioff, Phys. Rev {\bf D}34, 
3831 (1986)

\bibitem{jackiw} R. Balian and R. Veneroni, Phys. Rev. Lett.
{\bf 47}, 1353 (1981)

\bibitem{chetcope}G.J. Cheetham and E.J. Copeland, Phys. Rev. {\bf D}53, 4125 
(1996)

\bibitem{feit} M. Feit, J. A. Fleck Jr., and A. Steiger, J. Comp. Phys. {\bf 
47}, 412 (1982)
 

\bibitem{mean} J.P. Paz and F.D. Mazzitelli, Phys. Rev. {\bf D37}, 2170
(1988); F.D. Mazzitelli and J.P. Paz, Phys. Rev. {\bf D39}, 2234 (1989);
S.A. Ramsey and B.L. Hu, Phys. Rev. D56, 661 (1997);
D. Boyanovsky and H.J. de Vega,  Phys. Rev.{\bf D47}, 2343 (1993); M. Gleiser 
and R. Ramos, Phys. Rev {\bf D50}, 2441 (1994)

\bibitem{boya1} F. Cooper, S. Habib, Y. Kluger, E. Mottola, J.P. Paz, and P.
R. Anderson, Phys. Rev. {\bf D50}, 2848 (1994); F. Cooper, 
Y. Kluger, E. Mottola, and  J.P. Paz, Phys. Rev. {\bf D51}, 2377 (1995); 
D. Boyanovsky and H.J. de Vega, in Ref. \cite{mean}

\bibitem{boya2} D. Boyanovsky, H.J. de Vega, and R. Holman, Phys. Rev. 
{\bf D49}, 2769 (1994); D. Boyanovsky, H.J. de Vega, R. Holman, D.-S. Lee, 
and A. Singh, Phys. Rev. 
{\bf D51}, 4419 (1995); D. Boyanovsky, D. Cormier, H.J. de Vega, R. Holman, 
and S. Prem Kumar, Phys. Rev. {\bf D57}, 2166 (1998)

\bibitem{boya3} D. Boyanovsky, C. Destri, H.J. de Vega, R. Holman, and
 J. Salgado, Phys. Rev. {\bf D57}, 7388 (1998)

\bibitem{boya4} D. Boyanovsky, H.J. de Vega, and R. Holman, {\it 
Non-equilibrium phase transitions in condensed matter and cosmology: 
spinodal 
decomposition, condensates and defects}; in Lectures delivered at the NATO 
Advanced Study Institute: Topological Defects and the Non-Equilibrium Dynamics
of Symmetry Breaking Phase Transition; hep-ph/9909372

\bibitem{mihaila} B. Mihaila, J. Dawson, F. Cooper, M. Brewster, and S. Habib, 
{\it The quantum roll in d-dimensions and the large-d expansion}; 
hep-ph/9808234






\end{references}
\end{document}